\documentclass[a4paper,11pt]{article}
\usepackage{graphicx,epsfig,picins}
\usepackage{fancyhdr,fancybox}
\usepackage{indentfirst}
\usepackage{verbatim}
\usepackage[sort&compress, numbers]{natbib}
\usepackage{geometry}
\usepackage{extarrows} 
\usepackage[small]{caption2}
\usepackage{microtype}
\DisableLigatures[f]{encoding = *, family = * }

\usepackage{times}     

\usepackage{amssymb}                 
\usepackage{amsthm}
\usepackage{bbm}
\usepackage{hyperref}  

\newcommand{\paperfont}{\fontsize{12pt}{1.3\baselineskip}\selectfont}
\begin{document}

\theoremstyle{definition}
\makeatletter
\thm@headfont{\bf}
\makeatother
\newtheorem{definition}{Definition}
\newtheorem{example}{Example}
\newtheorem{theorem}{Theorem}
\newtheorem{lemma}{Lemma}
\newtheorem{corollary}{Corollary}
\newtheorem{remark}{Remark}
\newtheorem{proposition}{Proposition}

\lhead{}
\rhead{}
\lfoot{}
\rfoot{}

\renewcommand{\refname}{References}
\renewcommand{\figurename}{Figure}
\renewcommand{\tablename}{Table}
\renewcommand{\proofname}{Proof}

\newcommand{\diag}{\mathrm{diag}}
\newcommand{\one}{\mathbbm{1}}


\title{\textbf{Model simplification and loss of irreversibility}}
\author{Chen Jia$^{1,2}$ \\
\footnotesize $^1$Beijing Computational Science Research Center, Beijing 100094, P.R. China.\\
\footnotesize $^2$Department of Mathematical Sciences, The University of Texas at Dallas, Richardson, Taxes 75080, U.S.A.\\
\footnotesize Email: jiac@utdallas.edu}
\date{}                              
\maketitle                           
\thispagestyle{empty}                

\paperfont

\begin{abstract}
In this paper, we reveal a general relationship between model simplification and irreversibility based on the model of continuous-time Markov chains with time-scale separation. According to the topological structure of the fast process, we divide the states of the chain into the transient states and the recurrent states. We show that a two-time-scale chain can be simplified to a reduced chain in two different ways: removal of the transient states and aggregation of the recurrent states. Both the two operations will lead to a decrease in the entropy production rate and its adiabatic part and will keep its non-adiabatic part the same. This suggests that although model simplification can retain almost all the dynamic information of the chain, it will lose some thermodynamic information as a trade-off. \\

\end{abstract}

\section{Introduction}
In the recent two decades, significant progresses have been made in the field of mesoscopic stochastic nonequilibrium thermodynamics \cite{jarzynski2011equalities, seifert2012stochastic, van2015ensemble}. The dynamic foundation of this theory turns out to be Markov processes which have a dual representation in terms of their probability distributions and trajectories. In the theory of stochastic thermodynamics, an equilibrium state is defined as a steady-state process with detailed balance and the deviation from the equilibrium state is usually characterized by the concept of the entropy production rate \cite{jiang2004mathematical, kjelstrup2008non, zhang2012stochastic}. When an open system is driven with a sustained energy supply from the environment, it will approach a nonequilibrium steady state with concomitant entropy production.

Recently, it has been shown that the entropy production rate consists of two nonnegative contributing terms: the adiabatic and non-adiabatic parts of the entropy production rates \cite{ge2010physical, esposito2010three}. The adiabatic part is also known as the housekeeping heat \cite{oono1998steady, hatano2001steady} and the non-adiabatic part is also referred to as the free energy dissipation rate. This decomposition is important because it reveals two different sources of irreversibility. The adiabatic part describes the irreversibility due to the breaking of detailed balance and the non-adiabatic part describes that due to the deviation from the steady state \cite{ge2010physical, esposito2010three}.

The underdamped Langevin equation provides a fundamental description of the motion of a Brownian particle. In the low Reynold number regime, the underdamped Langevin equation can be simplified to the overdamped one by elimination of the velocity variable. In recent years, it has been found that the underdamped and overdamped Langevin equations provide different values of the entropy production rate when a Brownian particle is placed in an inhomogeneous temperature field \cite{hondou2000unattainability, celani2012anomalous}. Since then, there has been several research focusing on the effects of coarse graining on the entropy production rate based on different coarse-graining procedures and different models \cite{rahav2007fluctuation, puglisi2010entropy, kawaguchi2013fluctuation, bo2014entropy, nakayama2015invariance, sohn2016critical}. From the mathematical perspective, the simplification of Markov models has several different mechanisms. This raises a natural question of whether there exists a general relationship between model simplification and irreversibility.

The master equation, which describes the dynamics of a continuous-time Markov chain, plays a fundamental role in the field of stochastic thermodynamics not only because it provides an effective way to model various living systems, but also because any Markov process can generally be approximated by a continuous-time Markov chain. In practice, it often occurs that the state transitions of a chain possess two separate time scales. In this case, it is possible to simplify the original chain to a reduced one that will lose very little dynamic information of the original chain. In fact, the coarse graining of two-time-scale Markov chains have been studied by physicists and engineers for a long time \cite{haken1983synergetics, bobbio1986aggregation, reibman1989analysis, bobbio1990computing, malhotra1994stiffness, stewart1994introduction, gorban2006model, pigolotti2008coarse, ullah2012simplification} and its relationship with stochastic thermodynamics is also considered \cite{santillan2011irreversible, esposito2012stochastic, gaveau2014relative}. Recently, the effectiveness of the simplification approach has been established rigourously from a mathematical perspective \cite{yin2013continuous, jia2016reduction}. Thus continuous-time Markov chains with time-scale separation provide us a ideal model to study the relationship between model simplification and irreversibility.

In this paper, we perform a general analysis on the connection between model simplification and irreversibility based on the model of two-time-scale Markov chains. According to the topological structure of the fast process, we divide the states of the chain into the transient states and the recurrent states. We reveal that a two-time-scale Markov chain can be simplified to a reduced chain in two different ways: removal of the transient states and aggregation of the recurrent states, where the transient states can be removed via a series of transformations similar to the $Y-\Delta$ transformations in the circuit theory \cite{akers1960use}. It turns out that both the two operations will lead to a decrease in the entropy production rate and its adiabatic part and will keep its non-adiabatic part the same. This suggests that although model simplification can retain almost all the dynamic information of the chain, it must lose some thermodynamic information as a trade-off.

\section{Model}
In this paper, we consider a molecular system modeled by a continuous-time Markov chain, also called Markov jump process, $X = \{X_t:t\geq 0\}$ with state space $S = \{1,\cdots,n\}$ and generator matrix $Q = (q_{ij})$, where $q_{ij}$ with $i\neq j$ denotes the transition rate from state $i$ to $j$ and $q_{ii} = -\sum_{j\neq i}q_{ij}$. Let $p(t) = (p_1(t),\cdots,p_n(t))$ denote the probability distribution of the chain. Then the dynamics of the chain is governed by the master equation
\begin{equation}\label{master}
\frac{dp(t)}{dt} = p(t)Q.
\end{equation}
Without loss of generality, we assume that the chain is irreducible, which means that for any states $i$ and $j$, there are a sequence of states $i_1,i_2,\cdots,i_k$, such that $i_1=i$, $i_k=j$, and
\begin{equation*}
q_{i_1i_2}q_{i_2i_3}\cdots q_{i_{k-1}i_k}>0.
\end{equation*}
If the chain is irreducible, it has a unique steady-state distribution $p^{ss} = (p^{ss}_1,\cdots,p^{ss}_n)$ which satisfies $p^{ss}Q = 0$ \cite{norris1998markov}. Recall that the free energy $F(t)$ of the chain is usually defined as the relative entropy between the probability distribution $p(t)$ and the steady-state distribution $p^{ss}$ \cite{ge2010physical}:
\begin{equation*}
F(t) = \sum_{i=1}^np_i(t)\log\frac{p_i(t)}{p^{ss}_i}.
\end{equation*}

Let $J_{ij}(t) = p_i(t)q_{ij}$ denote the probability flux from state $i$ to $j$ and let $J^{ss}_{ij} = p^{ss}_iq_{ij}$ denote the corresponding steady-state flux. Recall that the entropy production rate $e_p(t)$ of the chain is defined as
\begin{equation*}
e_p(t) = \sum_{i,j=1}^nJ_{ij}(t)\log\frac{J_{ij}(t)}{J_{ji}(t)} \geq 0.
\end{equation*}
Recently, it has been shown that the entropy production rate $e_p(t)$ can be decomposed into the sum of two nonnegative terms:
\begin{equation}
e_p(t) = e_p^{(ad)}(t)+e_p^{(na)}(t),
\end{equation}
where $e_p^{(ad)}(t)$ and $e_p^{(na)}(t)$ are the adiabatic and non-adiabatic parts of the entropy production rate, respectively \cite{ge2010physical, esposito2010three}. Specifically, the adiabatic part $e_p^{(ad)}(t)$, also known as the housekeeping heat, is defined as
\begin{equation*}
e_p^{(ad)}(t) = \sum_{i,j=1}^nJ_{ij}(t)\log\frac{J^{ss}_{ij}}{J^{ss}_{ji}} \geq 0
\end{equation*}
and the non-adiabatic part $e_p^{(na)}(t)$, also called the free energy dissipation rate, is defined as
\begin{equation*}
e_p^{(na)}(t) = \sum_{i,j=1}^nJ_{ij}(t)\log\frac{p_i(t)p^{ss}_j}{p_j(t)p^{ss}_i} \geq 0.
\end{equation*}
It is easy to check that the non-adiabatic entropy production rate is exactly the dissipation rate of the free energy $F(t)$, that is,
\begin{equation*}
e_p^{(na)}(t) = -\frac{dF(t)}{dt}.
\end{equation*}

It is well-known that the entropy production rate $e_p(t)$ characterizes the irreversibility of a nonequilibrium process. In fact, there are two different sources of irreversibility. The adiabatic part $e_p^{(ad)}(t)$ describes the irreversibility due to the breaking of detailed balance and the non-adiabatic part $e_p^{(na)}(t)$ describes that due to the deviation from the steady state \cite{ge2010physical, esposito2010three}.

\section{Simplification of two-time-scale Markov chains}
In applications, it often occurs that some transition rates of a Markov chain are much faster than other ones. In this case, the transition rates of the chain have two separate time scales. To be specific, we assume that there is a threshold value $\lambda$ for the transition rates. If $q_{ij}$ is larger than the threshold value $\lambda$, then it is called a fast transition rate. Otherwise, it is called a slow transition rate. Intuitively, if we discard the process with fast time scale, then the original chain may be simplified.

In fact, the simplification of two-time-scale Markov chains has bee studied by physicists, engineers, and mathematicians for a long time \cite{bobbio1986aggregation, reibman1989analysis, bobbio1990computing, malhotra1994stiffness, stewart1994introduction, gorban2006model, pigolotti2008coarse, yin2013continuous, jia2016reduction}. Here we would like to review the simplification approach of two-time-scale chains based on the method of averaging. For convenience, we introduce several notations. If $i\neq j$, we define
\begin{equation*}
\begin{cases}
q^f_{ij} = q_{ij}, q^s_{ij} = 0, &\textrm{if $q_{ij}$ is a fast transition rate}, \\
q^f_{ij} = 0, q^s_{ij} = q_{ij}, &\textrm{if $q_{ij}$ is a slow transition rate}.
\end{cases}
\end{equation*}
In addition, we set $q^f_{ii} = -\sum_{j\neq i}q^f_{ij}$ and $q^s_{ii} = -\sum_{j\neq i}q^s_{ij}$. Then the generator matrix $Q$ can be rewritten as
\begin{equation*}
Q = Q^f+Q^s,
\end{equation*}
where $Q^f = (q^f_{ij})$ is a generator matrix that governs the fast process of the chain and $Q^s = (q^s_{ij})$ is another generator matrix that governs the slow process of the chain.

To proceed, we focus on the fast process governed by the generator matrix $Q^f$. Let $i$ and $j$ be two arbitrary states. Then we say that $i$ leads to $j$ and write $i\rightarrow j$, if there exist a sequence of states $i_1,\cdots,i_k$, such that $i_1 = i$, $i_k = j$, and
\begin{equation*}
q^f_{i_1i_2}q^f_{i_2i_3}\cdots q^f_{i_{k-1}i_k} > 0.
\end{equation*}
In addition, we say that $i$ communicates with $j$ and write $i\leftrightarrow j$ if $i\rightarrow j$ and $j\rightarrow i$. It is easy to check that $\leftrightarrow$ is an equivalence relation on the state space $S$. Thus $S$ can be decomposed into the union of different communicating classes. Although the original chain is irreducible, its fast process may possess multiple communicating classes.

In general, there are two types of states that must be distinguished. Let $i$ be a state in the communicating class $C$. If there exists a state $j\notin C$ such that $i\rightarrow j$, then $i$ is called a transient state. Otherwise, $i$ is called a recurrent state. It is easy to check that if $i$ is a recurrent state, then all the states in $C$ are recurrent \cite{norris1998markov}. In the case, the communicating class $C$ is called a recurrent class.

\begin{figure}[!htb]
\begin{center}
\centerline{\includegraphics[width=0.5\textwidth]{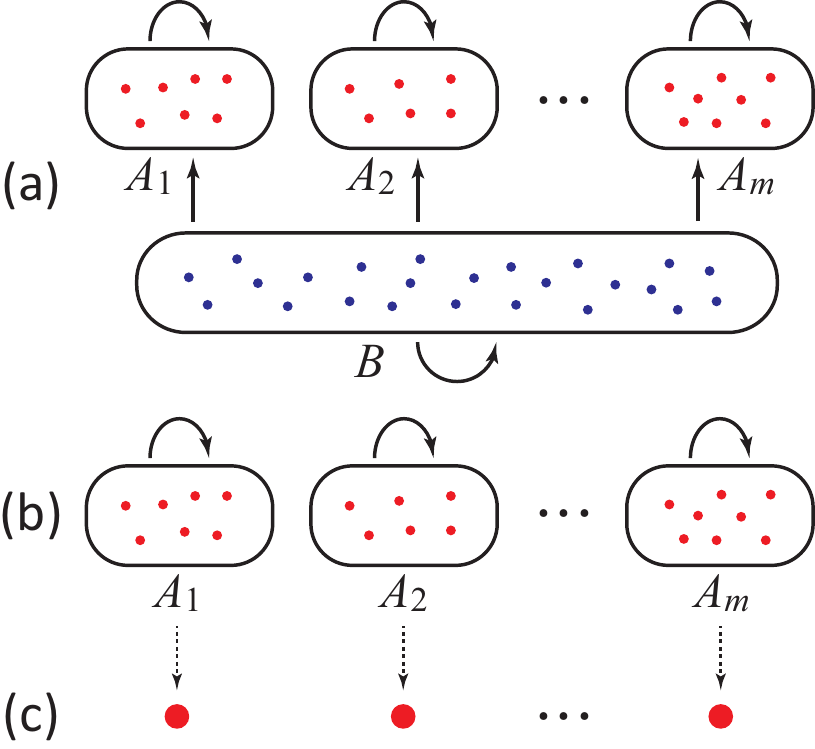}}
\caption{\textbf{The simplification of two-time-scale Markov chains.} (a) The canonical decomposition of the state space. According to the topological structure of the fast process, the state space $S$ can be decomposed into the union of multiple recurrent classes $A_1,\cdots,A_m$ and the collection $B$ of all transient states. (b) The removal of the transient states of the fast process. (c) The aggregation of the recurrent states of the fast process.}\label{simplification}
\end{center}
\end{figure}

Let $A$ denote the collection of all the recurrent states of the fast process and let $B$ denote that of all the transient states of the fast process. Then $A$ can be decomposed into the union of different recurrent classes $A_1,\cdots,A_m$ and thus the state space $S$ can be decomposed into
\begin{equation*}
S = A\cup B = A_1\cup\cdots\cup A_m\cup B,
\end{equation*}
as illustrated in Figure \ref{simplification}(a).
In this way, the generator matrix $Q^f$ can be represented as a block matrix
\begin{equation*}
Q^f = \begin{pmatrix}
Q^f_{11} &        &          &                \\
         & \ddots &          &                \\
         &        & Q^f_{mm} &                \\
Q^f_{B1} & \cdots & Q^f_{Bm} & Q^f_{BB}
\end{pmatrix},
\end{equation*}
where $Q^f_{11},\cdots,Q^f_{mm}$ are $m$ generator matrices that govern the transitions within the recurrent classes $A_1,\cdots,A_m$, respectively. Similarly, we represent the generator matrices $Q$ and $Q^s$ as
\begin{equation*}
Q = \begin{pmatrix}
Q_{11} & \cdots & Q_{1m} & Q_{1B} \\
Q_{21} & \cdots & Q_{2m} & Q_{2B} \\
       & \cdots & \cdots &        \\
Q_{B1} & \cdots & Q_{Bm} & Q_{BB}
\end{pmatrix},\,\,\,
Q^s = \begin{pmatrix}
Q^s_{11} & \cdots & Q^s_{1m} & Q^s_{1B} \\
Q^s_{21} & \cdots & Q^s_{2m} & Q^s_{2B} \\
         & \cdots & \cdots   &          \\
Q^s_{B1} & \cdots & Q^s_{Bm} & Q^s_{BB}
\end{pmatrix},
\end{equation*}
and represent the probability distribution $p(t)$ as $p(t) = (p^1(t),\cdots,p^m(t),p^B(t))$. With these notations, the master equation \eqref{master} can be rewritten as
\begin{equation}\label{master2}\left\{
\begin{split}
\dot{p}^1(t) &= p^1(t)Q_{11}+\cdots+p^m(t)Q_{m1}+p^B(t)Q_{B1}, \\
&\cdots \\
\dot{p}^m(t) &= p^1(t)Q_{1m}+\cdots+p^m(t)Q_{mm}+p^B(t)Q_{Bm}, \\
\dot{p}^B(t) &= p^1(t)Q_{1B}+\cdots+p^m(t)Q_{mB}+p^B(t)Q_{BB}.
\end{split}\right.
\end{equation}

Let $\one$ denote the column vector whose components are all 1. Let $A_k$ be a recurrent class of the fast process and set
\begin{equation}\label{aggregation}
\pi_k(t) = p^k(t)\one = \sum_{i\in A_k}p_i(t).
\end{equation}
Then $\pi_k(t)$ denotes the probability that the chain stays in $A_k$ at time $t$. In addition, let $\mu^k$ denote the steady-state distribution of the generator matrix $Q^f_{kk}$. Since the transitions within $A_k$ are very fast, the states in $A_k$ will reach a quasi-steady state. This suggests that
\begin{equation}\label{quasi}
p^k(t)\approx \pi_k(t)\mu^k,\;\;\;k = 1,\cdots,m.
\end{equation}
Since $Q^f_{11},\cdots,Q^f_{mm}$ are generator matrices, we have $Q^f_{11}\one = \cdots = Q^f_{mm}\one = 0$. From \eqref{master2} and \eqref{quasi}, we obtain that
\begin{equation}\label{master3}\left\{
\begin{split}
\dot\pi_1(t) &\approx \pi_1(t)\mu^1Q^s_{11}\one+\cdots+\pi_m(t)\mu^mQ^s_{m1}\one+p^B(t)Q_{B1}\one, \\
&\cdots \\
\dot\pi_m(t) &\approx \pi_1(t)\mu^1Q^s_{1m}\one+\cdots+\pi_m(t)\mu^mQ^s_{mm}\one+p^B(t)Q_{Bm}\one, \\
\dot{p}^B(t) &\approx \pi_1(t)\mu^1Q_{1B}+\cdots+\pi_m(t)\mu^mQ_{mB}+p^B(t)Q_{BB}.
\end{split}\right.
\end{equation}
In this equation, $\pi_1(t),\cdots,\pi_m(t)$ are $m$ slow variables and $p^B(t)$ is a fast variable. If we focus on the fast variable $p^B(t)$, then the slow variables $\pi_1(t),\cdots,\pi_m(t)$ can be frozen. From \eqref{master3}, the steady-state value $p^B(t)_{ss}$ of the fast variable is approximately given by
\begin{equation}\label{steady}
p^B(t)_{ss}\approx -(\pi_1(t)\mu^1Q_{1B}+\cdots+\pi_m(t)\mu^mQ_{mB})Q_{BB}^{-1}.
\end{equation}
In addition, if we focus on the slow variables $\pi_1(t),\cdots,\pi_m(t)$, then we can think that the fast variable $p^B(t)$ has reached its steady-state value. From \eqref{master3} and \eqref{steady}, the dynamics of the slow variables $\pi_1(t),\cdots,\pi_m(t)$ is approximately governed by
\begin{equation}\label{master4}\left\{
\begin{split}
\dot\pi_1 &\approx \pi_1\mu^1(Q_{11}-Q_{1B}Q_{BB}^{-1}Q_{B1})\one + \cdots + \pi_m\mu^m(Q_{m1}-Q_{mB}Q_{BB}^{-1}Q_{B1})\one, \\
&\cdots \\
\dot\pi_m &\approx \pi_1\mu^1(Q_{1m}-Q_{1B}Q_{BB}^{-1}Q_{Bm})\one + \cdots + \pi_m\mu^m(Q_{mm}-Q_{mB}Q_{BB}^{-1}Q_{Bm})\one.
\end{split}\right.
\end{equation}

We make a crucial observation that the above equation is an approximate master equation. This suggests that the original chain can be simplified to a reduced chain with $m$ different states, each of which corresponds to a recurrent class of the fast process. Recall that we have decomposed the state space $S$ into the union of $A$ and $B$. Thus the generator matrix $Q$ can be represented as
\begin{equation*}
Q = \begin{pmatrix}
Q_{AA} & Q_{AB} \\
Q_{BA} & Q_{BB}
\end{pmatrix}.
\end{equation*}
Let $\hat{p}(t) = (\hat{p}_1(t),\cdots,\hat{p}_m(t))$ denote the probability distribution of the reduced chain.
From \eqref{master4}, the dynamics of the reduced chain is given by
\begin{equation*}
\frac{d\hat{p}(t)}{dt} = \hat{p}(t)\hat{Q},
\end{equation*}
where
\begin{equation}
\hat{Q} =
\begin{pmatrix}
\mu^1 &        &       \\
      & \ddots &       \\
      &        & \mu^m
\end{pmatrix}(Q_{AA}-Q_{AB}Q_{BB}^{-1}Q_{BA})
\begin{pmatrix}
\one &        &       \\
     & \ddots &       \\
     &        & \one
\end{pmatrix}
\end{equation}
is the generator matrix of the reduced chain.

\section{Two steps of model simplification}
The above analysis suggests that the simplification of a two-time-scale Markov chain can be decomposed into two steps. In the first step, the original chain is simplified to a reduced chain by removal of the transient states of the fast process, as illustrated in Figure \ref{simplification}(b). In this way, the state space $S$ is reduced to the state space $A$ and the generator matrix $Q$ is reduced to
\begin{equation*}
\tilde{Q} = Q_{AA}-Q_{AB}Q_{BB}^{-1}Q_{BA}.
\end{equation*}
In the second step, the original chain is further simplified to a reduced chain by aggregation of the recurrent states of the fast process. To be specific, the states in each recurrent class will be aggregated into one state and thus the state space of the reduced chain can be viewed as the collection of all the recurrent classes, as illustrated in Figure \ref{simplification}(c). In this way, the state space $A$ is further reduced to the state space $\hat{S} = \{A_1,\cdots,A_m\}$ and the generator matrix $\tilde{Q}$ is further reduced to
\begin{equation}\label{second}
\hat{Q} = \begin{pmatrix}
\mu^1 &        &       \\
      & \ddots &       \\
      &        & \mu^m
\end{pmatrix}\tilde{Q}
\begin{pmatrix}
\one &        &       \\
     & \ddots &       \\
     &        & \one
\end{pmatrix}.
\end{equation}

In the previous work, many authors have studied the coarse graining of a master equation with fast and slow variables and its relationship with stochastic thermodynamics. Among these studies, there is a popular model which considers a molecular system with $m$ mesostates, each of which contains $r$ microstates \cite{haken1983synergetics, santillan2011irreversible, esposito2012stochastic, kawaguchi2013fluctuation, bo2014entropy, nakayama2015invariance}. Thus the molecular system can be modeled by a continuous-time Markov chain whose state space can be represented by a two-dimensional lattice
\begin{equation*}
S = \{(x,y): x = 1,\cdots m, y = 1,\cdots,r\},
\end{equation*}
where $x$ represents a mesostate and $(x,y)$ represents a microstate. The model further assumes that the the transitions between microstates belonging to the same mesostate are much faster than those belonging to different mesostates, which is often called the adiabatic hypothesis \cite{haken1983synergetics}. Under this assumption, $x$ is a slow variable and $y$ is fast variable. If we plot the state space $S$ as a two-dimensional lattice, then the vertical transition rates are much faster than the horizontal ones.

Since the vertical transition rates are fast and the horizontal ones are slow, it is easy to see that all microstates are recurrent states of the fast process and the model has $m$ recurrent classes. Each recurrent class corresponds to a mesostate and the microstates belonging to a mesostate can be aggregated. Based on the so-called adiabatic approximation, it has been shown in the previous papers \cite{santillan2011irreversible, esposito2012stochastic} that the original model can be simplified to a coarse-grained one which only considers the slow variable $x$, whose dynamics is described by the following master equation:
\begin{equation*}
\frac{d\hat{p}(t)}{dt} = \hat{p}(t)\hat{Q},
\end{equation*}
where $\hat{Q} = (\hat{q}_{x,x'})$ is the transition rate matrix of the coarse-grained model with
\begin{equation*}
\hat{q}_{x,x'} = \sum_{y,y'}p^{ss}(y|x)q_{(x,y),(x',y')},\;\;\;x\neq x'.
\end{equation*}
Here, $p^{ss}(y|x)$ is the steady-state probability of the fast variable $y$ when the slow variable $x$ is frozen. In fact, that the above formula is essentially the same as \eqref{second}. This suggests that the above coarse-graining procedure is essentially the aggregation of the recurrent states of the fast process.

In addition, there are also some papers developing another coarse-graining procedure of master equations, which turns out to be equivalent to the removal of the transient states of the fast process \cite{pigolotti2008coarse, puglisi2010entropy, ullah2012simplification, jia2016reduction}. In this paper, we present a unified treatment of the above two types of coarse-graining procedures.

\section{Removal of transient states}
The first step of model simplification is to remove the transient states of the fast process. In this way, the original chain can be simplified to a reduced chain with state space $A$ and generator matrix
\begin{equation}\label{effective}
\tilde{Q} = (\tilde{q}_{ij}) = Q_{AA}-Q_{AB}Q_{BB}^{-1}Q_{BA}.
\end{equation}
Assume that the system has $n_A$ recurrent states. Let $\tilde{p}(t) = (\tilde{p}_1(t),\cdots,\tilde{p}_{n_A}(t))$ denote the probability distribution of the reduced chain and
let $\tilde{p}^{ss}(t) = (\tilde{p}^{ss}_1(t),\cdots,\tilde{p}^{ss}_{n_A}(t))$ denote the corresponding steady-state distribution. In addition, let $\tilde{J}_{ij}(t) = \tilde{p}_i(t)\tilde{q}_{ij}$ denote the probability flux of the reduced chain from state $i$ to $j$ and let $\tilde{J}^{ss}_{ij}(t) = \tilde{p}^{ss}_i(t)\tilde{q}_{ij}$ denote the corresponding steady-state flux.

In fact, the probability fluxes of the reduced chain can also be obtained from those of the original chain by a series of transformations. To see this, we first consider the case where there is only one transient state, which is assumed to be state $n$. From \eqref{effective}, the transition rate of the reduced chain is given by
\begin{equation}\label{simplerate}
\tilde{q}_{ij} = q_{ij}-q_{in}q_{nj}/q_{nn},\;\;\;i,j = 1,\cdots,n-1.
\end{equation}
This shows that the transition rate $q_{ij}$ of the original chain and the transition rate $\tilde{q}_{ij}$ of the reduced chain differ by a term $-q_{in}q_{nj}/q_{nn}$, where $q_{in}$ is the transition rate from the recurrent state $i$ to the transient state $n$ and the quantity $-q_{nj}/q_{nn}$ is the transition probability of the embedded chain from the transient state $n$ to the recurrent state $j$.

Since the dynamics of the original chain and that of the reduced chain are closed to each other, we have \cite{jia2016reduction}
\begin{equation}\label{close}
\begin{cases}
\tilde{p}_i(t)\approx p_i(t), &\textrm{if $i$ is a recurrent state}, \\
\tilde{p}_i(t)\approx 0, &\textrm{if $i$ is a transient state}.
\end{cases}
\end{equation}
\footnote{From the mathematical perspective, if the initial distribution of the original chain concentrates on the state space $A$, then the probability distributions of the original and reduced chains will agree with each other over the whole time axis. However, if the initial distribution does not concentrates on the state space $A$, then model simplification may cause large errors within a very short time, but the probability distributions of the two chains will coincide with each other afterwards \cite{jia2016reduction}.}
From \eqref{simplerate}, the probability flux of the reduced chain is approximately given by
\begin{equation}\label{simple}
\tilde{J}_{ij}(t)\approx J_{ij}(t)-J_{in}(t)J_{nj}(t)/J_{nn}(t),\;\;\;i,j = 1,\cdots,n-1.
\end{equation}
Thus the steady-state flux of the reduced chain is approximately given by
\begin{equation}
\tilde{J}^{ss}_{ij}\approx J^{ss}_{ij}-J^{ss}_{in}J^{ss}_{nj}/J^{ss}_{nn},\;\;\;i,j = 1,\cdots,n-1.
\end{equation}

Since state $n$ is a transient state of the fast process, it must reach a quasi-steady state. This suggests that the the sum of the probability fluxes flowing into state $n$ and that flowing out of state $n$ should be approximately the same, that is,
\begin{equation}\label{flow}
\sum_{i=1}^{n-1}J_{in}(t) \approx \sum_{j=1}^{n-1}J_{nj}(t) = -J_{nn}(t).
\end{equation}
Let us recall the following important inequality which is called the log sum inequality in the literature \cite{cover2006elements}: for nonnegative real numbers $a_1,\cdots,a_n$ and $b_1,\cdots,b_n$, we have
\begin{equation*}
\left(\sum_{i=1}^na_i\right)\log\frac{\sum_{i=1}^na_i}{\sum_{i=1}^nb_i} \leq \sum_{i=1}^na_i\log\frac{a_i}{b_i}.
\end{equation*}
By the log sum inequality and \eqref{flow}, the entropy production rate of the reduced chain satisfies
\begin{equation*}
\begin{split}
\tilde{e}_p &= \sum_{i,j=1}^{n-1}\tilde{J}_{ij}\log\frac{\tilde{J}_{ij}}{\tilde{J}_{ji}}
\approx \sum_{i,j=1}^{n-1}(J_{ij}-J_{in}J_{nj}/J_{nn})
\log\frac{J_{ij}-J_{in}J_{nj}/J_{nn}}{J_{ji}-J_{jn}J_{ni}/J_{nn}} \\
&\leq \sum_{i,j=1}^{n-1}(J_{ij}\log\frac{J_{ij}}{J_{ji}} - J_{in}J_{nj}/J_{nn}\log\frac{J_{in}J_{nj}/J_{nn}}{J_{jn}J_{ni}/J_{nn}}) \\
&= \sum_{i,j=1}^{n-1}J_{ij}\log\frac{J_{ij}}{J_{ji}} - \sum_{i,j=1}^{n-1}J_{in}J_{nj}/J_{nn}\log\frac{J_{in}}{J_{ni}} - \sum_{i,j=1}^{n-1}J_{in}J_{nj}/J_{nn}\log\frac{J_{nj}}{J_{jn}} \\
&= \sum_{i,j=1}^{n-1}J_{ij}\log\frac{J_{ij}}{J_{ji}} + \sum_{i=1}^{n-1}J_{in}\log\frac{J_{in}}{J_{ni}} + \sum_{j=1}^{n-1}J_{nj}\log\frac{J_{nj}}{J_{jn}} \\
&= \sum_{i,j=1}^nJ_{ij}\log\frac{J_{ij}}{J_{ji}} = e_p.
\end{split}
\end{equation*}
and the adiabatic entropy production rate of the reduced chain satisfies
\begin{equation*}
\begin{split}
\tilde{e}_p^{(ad)} &= \sum_{i,j=1}^{n-1}\tilde{J}_{ij}\log\frac{\tilde{J}^{ss}_{ij}}{\tilde{J}^{ss}_{ji}}
= \sum_{i,j=1}^{n-1}\frac{\tilde{p}_i}{\tilde{p}^{ss}_i}\tilde{J}^{ss}_{ij}
\log\frac{\tilde{J}^{ss}_{ij}}{\tilde{J}^{ss}_{ji}} \\
&\approx \sum_{i,j=1}^{n-1}\frac{p_i}{p^{ss}_i}(J^{ss}_{ij}-J^{ss}_{in}J^{ss}_{nj}/J^{ss}_{nn})
\log\frac{J^{ss}_{ij}-J^{ss}_{in}J^{ss}_{nj}/J^{ss}_{nn}}{J^{ss}_{ji}-J^{ss}_{jn}J^{ss}_{ni}/J^{ss}_{nn}} \\
&\leq \sum_{i,j=1}^{n-1}\frac{p_i}{p^{ss}_i}(J^{ss}_{ij}\log\frac{J^{ss}_{ij}}{J^{ss}_{ji}} - J^{ss}_{in}J^{ss}_{nj}/J^{ss}_{nn}\log\frac{J^{ss}_{in}J^{ss}_{nj}/J^{ss}_{nn}}{J^{ss}_{jn}J^{ss}_{ni}/J^{ss}_{nn}}) \\
&= \sum_{i,j=1}^{n-1}J_{ij}\log\frac{J^{ss}_{ij}}{J^{ss}_{ji}} - \sum_{i,j=1}^{n-1}J_{in}J_{nj}/J_{nn}\log\frac{J^{ss}_{in}}{J^{ss}_{ni}} - \sum_{i,j=1}^{n-1}J_{in}J_{nj}/J_{nn}\log\frac{J^{ss}_{nj}}{J^{ss}_{jn}} \\
&= \sum_{i,j=1}^{n-1}J_{ij}\log\frac{J^{ss}_{ij}}{J^{ss}_{ji}} + \sum_{i=1}^{n-1}J_{in}\log\frac{J^{ss}_{in}}{J^{ss}_{ni}} + \sum_{j=1}^{n-1}J_{nj}\log\frac{J^{ss}_{nj}}{J^{ss}_{jn}} \\
&= \sum_{i,j=1}^nJ_{ij}\log\frac{J^{ss}_{ij}}{J^{ss}_{ji}} = e_p^{(ad)}.
\end{split}
\end{equation*}
This shows that the entropy production rate and its adiabatic part must decrease after removal of the transient state $n$. In addition, from \eqref{close} and \eqref{flow}, the free energy of the reduced chain satisfies
\begin{equation*}
\tilde{F} = \sum_{i=1}^{n-1}\tilde{p}_i\log\frac{\tilde{p}_i}{\tilde{p}^{ss}_i}
\approx \sum_{i=1}^np_i\log\frac{p_i}{p^{ss}_i} = F
\end{equation*}
and the non-adiabatic entropy production rate of the reduced chain satisfies
\begin{equation*}
\begin{split}
\tilde{e}_p^{(na)}
&= \sum_{i,j=1}^{n-1}\tilde{J}_{ij}\log\frac{\tilde{p}_i\tilde{p}^{ss}_j}{\tilde{p}_j\tilde{p}^{ss}_i}
\approx \sum_{i,j=1}^{n-1}(J_{ij}-J_{in}J_{nj}/J_{nn})\log\frac{p_ip^{ss}_j}{p_jp^{ss}_i} \\
&= \sum_{i,j=1}^{n-1}J_{ij}\log\frac{p_ip^{ss}_j}{p_jp^{ss}_i} - \sum_{i,j=1}^{n-1}J_{in}J_{nj}/J_{nn}\log\frac{p_i}{p^{ss}_i} - \sum_{i,j=1}^{n-1}J_{in}J_{nj}/J_{nn}\log\frac{p^{ss}_j}{p_j} \\
&= \sum_{i,j=1}^{n-1}J_{ij}\log\frac{p_ip^{ss}_j}{p_jp^{ss}_i} + \sum_{i=1}^{n-1}J_{in}\log\frac{p_i}{p^{ss}_i} + \sum_{j=1}^{n-1}J_{nj}\log\frac{p^{ss}_j}{p_j} \\
&\approx \sum_{i,j=1}^{n-1}J_{ij}\log\frac{p_ip^{ss}_j}{p_jp^{ss}_i} + \sum_{i=1}^{n-1}J_{in}\log\frac{p_ip^{ss}_n}{p_np^{ss}_i} + \sum_{j=1}^{n-1}J_{nj}\log\frac{p_np^{ss}_j}{p_jp^{ss}_n} \\
&= \sum_{i,j=1}^nJ_{ij}\log\frac{p_ip^{ss}_j}{p_jp^{ss}_i} = e_p^{(na)}.
\end{split}
\end{equation*}
This shows that the free energy and the non-adiabatic entropy production rate will remain the same after removal of the transient state $n$.

We next deal with the general case where there are $n_A$ transient states, which are assumed to be states $n-n_A+1,\cdots,n$. In this case, the probability fluxes of the reduced chain can be obtained by carrying out the transformation \eqref{simple} successively, which in essence remove the transient states one by one. To be specific, the series of transformations can be defined inductively as follows: set
\begin{equation*}
J^0_{ij}(t) = J_{ij}(t),\;\;\;i,j = 1,2,\cdots,n.
\end{equation*}
and for any $k = 1,\cdots,n_A$, set
\begin{equation}\label{transformation}
J^k_{ij}(t) = J^{k-1}_{ij}(t)-J^{k-1}_{ii_k}(t)J^{k-1}_{i_kj}(t)/J^{k-1}_{i_ki_k}(t),\;\;\;i,j = 1,\cdots i_k-1,
\end{equation}
where $i_k = n-k+1$ indexes the transient state to be removed in the $k$th step and $J^k_{ij}(t)$ denotes the probability flux from state $i$ to $j$ after the transient states $i_1,\cdots,i_k$ are removed. After all transient states are removed, we obtain the probability flux of the reduced chain:
\begin{equation*}
\tilde{J}_{ij}(t)\approx J^{n_A}_{ij}(t).
\end{equation*}
The above analysis shows that the probability fluxes of the reduced chain can be obtained by removing the transient states one by one. Once a transient state is removed, the adiabatic entropy production rate and will decrease, while the free energy and the non-adiabatic entropy production rate will remain the same. Thus the total entropy production rate must decrease after all transient states are removed.

Before leaving this section, we point out that the series of transformations defined in \eqref{transformation} have almost the same form as the so-called $Y-\Delta$ transformations in the circuit theory \cite{akers1960use}. In fact, it has been found that a reversible Markov chain is equivalent to an RC circuit and the master equation of the Markov chain is formally equivalent to Kirchoff's law of the RC circuit \cite{doyle1984random, ullah2012simplification}. In an RC circuit, if some capacitors have much smaller capacitances than other capacitors, then the original circuit can be simplified to a reduced one by removal of those small capacitors via the $Y-\Delta$ transformations. Thus the simplification of an RC circuit by removal of the small capacitors is equivalent to the simplification of a reversible chain by removal of the transient states.

\section{Aggregation of recurrent states}
Since the transient states of the fast process have been removed in the first step of model simplification, we can assume that the fast process has only recurrent states. Let $A_1,\cdots,A_m$ denote all the recurrent classes of the fast process. The second step of model simplification is to aggregate the states in each recurrent class into one state. In this way, the original chain can be simplified to a reduced chain with state space $\hat{S} = \{A_1,\cdots,A_m\}$ and generator matrix
\begin{equation*}
\hat{Q} = (\hat{q}_{kl}) = \begin{pmatrix}
\mu^1 &        &       \\
      & \ddots &       \\
      &        & \mu^m
\end{pmatrix}Q
\begin{pmatrix}
\one &        &       \\
     & \ddots &       \\
     &        & \one
\end{pmatrix},
\end{equation*}
where
\begin{equation}\label{effective2}
\hat{q}_{kl} = \mu^kQ_{kl}\one,\;\;\;k,l=1,\cdots,m.
\end{equation}

Let $\hat{p}(t) = (\hat{p}_1(t),\cdots,\hat{p}_m(t))$ denote the probability distribution of the reduced chain and let $\hat{p}^{ss}(t) = (\hat{p}^{ss}_1(t),\cdots,\hat{p}^{ss}_m(t))$ denote the corresponding steady-state distribution. In addition, let $\hat{J}_{kl}(t) = \hat{p}_k(t)\hat{q}_{kl}$ denote the probability flux of the reduced chain from state $A_k$ to $A_l$ and let $\hat{J}^{ss}_{kl}(t) = \hat{p}^{ss}_k(t)\hat{q}_{kl}$ denote the corresponding steady-state flux. From \eqref{quasi} and \eqref{effective2}, the probability flux of the reduced chain is approximately given by
\begin{equation*}
\hat{J}_{kl}(t) = \hat{p}_k(t)\mu^kQ_{kl}\one \approx p^k(t)Q_{kl}\one = \sum_{i\in A_k,j\in A_l}p_i(t)q_{ij} = \sum_{i\in A_k,j\in A_l}J_{ij}(t).
\end{equation*}
If state $i$ is in the recurrent class $A_k$, it follows from \eqref{quasi} that
\begin{equation}\label{prob}
p_i(t)\approx \hat{p}_k(t)\mu_i,\;\;\;p^{ss}_i(t)\approx \hat{p}^{ss}_k(t)\mu_i.
\end{equation}
This shows that
\begin{equation}\label{flux}
J_{ij}(t)\approx \hat{p}_k(t)\mu_iq_{ij},\;\;\;J^{ss}_{ij}(t)\approx \hat{p}^{ss}_k(t)\mu_iq_{ij}.
\end{equation}
By the log sum inequality, the entropy production rate of the reduced chain satisfies
\begin{equation*}
\begin{split}
\hat{e}_p &= \sum_{k,l=1}^m\hat{J}_{kl}\log\frac{\hat{J}_{kl}}{\hat{J}_{lk}}
\approx \sum_{k,l=1}^m\sum_{i\in A_k,j\in A_l}J_{ij}
\log\frac{\sum_{i\in A_k,j\in A_l}J_{ij}}{\sum_{i\in A_k,j\in A_l}J_{ji}} \\
&\leq \sum_{k,l=1}^m\sum_{i\in A_k,j\in A_l}J_{ij}\log\frac{J_{ij}}{J_{ji}}
= \sum_{i,j=1}^nJ_{ij}\log\frac{J_{ij}}{J_{ji}} = e_p.
\end{split}
\end{equation*}
and from \eqref{flux}, the adiabatic entropy production rate of the reduced chain satisfies
\begin{equation*}
\begin{split}
\hat{e}_p^{(ad)} &= \sum_{k,l=1}^m\hat{J}_{kl}\log\frac{\hat{J}^{ss}_{kl}}{\hat{J}^{ss}_{lk}}
= \sum_{k,l=1}^m\frac{\hat{p}_k}{\hat{p}^{ss}_k}\hat{J}^{ss}_{kl}\log\frac{\hat{J}^{ss}_{kl}}{\hat{J}^{ss}_{lk}} \\
&\approx \sum_{k,l=1}^m\frac{\hat{p}_k}{\hat{p}^{ss}_k}\sum_{i\in A_k,j\in A_l}J^{ss}_{ij}
\log\frac{\sum_{i\in A_k,j\in A_l}J^{ss}_{ij}}{\sum_{i\in A_k,j\in A_l}J^{ss}_{ji}} \\
&\leq \sum_{k,l=1}^m\frac{\hat{p}_k}{\hat{p}^{ss}_k}\sum_{i\in A_k,j\in A_l}J^{ss}_{ij}\log\frac{J^{ss}_{ij}}{J^{ss}_{ji}}
= \sum_{i,j=1}^nJ_{ij}\log\frac{J^{ss}_{ij}}{J^{ss}_{ji}} = e_p^{(ad)}.
\end{split}
\end{equation*}
This shows that the entropy production rate and its adiabatic part must decrease after aggregation of the recurrent states. In addition, from \eqref{aggregation} and \eqref{prob}, the free energy of the reduced chain satisfies
\begin{equation*}
\hat{F} = \sum_{k=1}^m\hat{p}_k\log\frac{\hat{p}_k}{\hat{p}^{ss}_k}
\approx \sum_{k=1}^m\sum_{i\in A_k}p_i\log\frac{p_i}{p^{ss}_i}
= \sum_{i=1}^np_i\log\frac{p_i}{p^{ss}_i} = F
\end{equation*}
and the non-adiabatic entropy production rate of the reduced chain satisfies
\begin{equation*}
\begin{split}
\hat{e}_p^{(na)} &= \sum_{k,l=1}^m\hat{J}_{kl}\log\frac{\hat{p}_k\hat{p}^{ss}_l}{\hat{p}_l\hat{p}^{ss}_k} \approx \sum_{k,l=1}^m\sum_{i\in A_k,j\in A_l}J_{ij}\log\frac{p_ip^{ss}_j}{p_jp^{ss}_i} \\
&= \sum_{i,j=1}^nJ_{ij}\log\frac{p_ip^{ss}_j}{p_jp^{ss}_i} = e_p^{(na)}.
\end{split}
\end{equation*}
This shows that the free energy and the non-adiabatic entropy production rate will remain the same after aggregation of the recurrent states.

The above analysis shows that the adiabatic entropy production rate will decrease, while the free energy and the non-adiabatic entropy production rate will remain the same after both the two steps of model simplification. Thus the total entropy production rate must decrease after model simplification.

\section{Conclusions}
In this paper, we study the relationship between model simplification and irreversibility. The results of this paper apply to all two-time-scale Markov chains and thus are quite general. We show that a two-time-scale Markov chain consists of a fast process and a slow process. Although the original chain is irreducible, its fast process may possess multiple communicating classes. According to the topological structure of the fast process, the state space $S$ can be decomposed into the union of the recurrent classes $A_1,\cdots,A_m$ and the set $B$ which consists of all transient states.

Based on the method of averaging, we show that the simplification of a two-time-scale Markov chain can be decomposed into two steps. The first step is to remove the transient states of the fast process and the second step is to aggregate the recurrent states of the fast process. Both the two steps will lead to a decrease in the total entropy production rate and its adiabatic part, and will keep the free energy and the non-adiabatic entropy production rate the same. This shows that the irreversibility due to the breaking of detailed balance will be lost and that due to the deviation of the steady state will remain the same after model simplification. Overall, a two-time-scale Markov chain will always lose a part of or all the irreversibility after model simplification.

The above analysis provides a deep physical insight. From a transcendental perspective, we hope that the reduced chain can retain both the dynamical and thermodynamical information of the original chain. However, the results of this paper show that this is impossible. Although model simplification retains almost all the dynamic information of the chain, it will lose some thermodynamic information as a trade-off. This fact is similar to the uncertainty principle of a quantum particle, whose position and momentum cannot be measured in a precise way simultaneously.

\section*{Acknowledgements}
The author is grateful to S.W. Wang for stimulating discussions and the anonymous reviewers for their valuable suggestions which improve the quality of this paper to a large extent.

\setlength{\bibsep}{5pt}
\small\bibliographystyle{apsrev4-1}
\bibliography{reduction}
\end{document}